# On the multifaceted journey for the invention of epitaxial quantum dots

Authors: Emanuele Pelucchi[a,] *

[a] *Tyndall National Institute, University College Cork, Cork , Ireland*

*Corresponding author.*

*E-mail address: emanuele.pelucchi@tyndall.ie;*

*Abstract*

Epitaxial semiconductor quantum dots have been, in the last 40 years or so, at the center of the research effort of a large community. The focus being on "semiconductor physics and devices", in view of the broad applications and potential, e.g., for efficient temperature insensitive lasers at telecom wavelengths, or as "artificial atoms" for quantum information processing. Our manuscript aims at addressing, with an historical perspective, the specifics of (III-V) epitaxial quantum dot early developments (largely for light emitting) and subsequent years. We will not only highlight the variety of epitaxial structures and methods, but also, intentionally glancing a didactic approach, discuss aspects that are, in general, little acknowledged or debated in the present literature. The analyses will also naturally bring us to examine some of current challenges, in a field which, despite sensational achievements, is, remarkably, still far from being mature in its developments and applications.

*Main Text-----------------------------------------------------------------------------------------------------------*

## *1. Introduction*

Epitaxial semiconductor quantum dots (QDs) have been at the centre of many research efforts since their inception in the early eighties. To notice, this also means that the field is already around 40 years old, the exact age somehow depending on what one considers as the founding date (a nontrivial task as might become clear from our discussion). Since then, many successes in terms of device physics and control of their (quantum) properties can be listed.

The aim of this manuscript is not, specifically, to give a complete account of those accomplishments. In many ways, one can find in the current literature several high-quality review papers, which do a very good job at that, and we will mention here a few recent ones, which can be a good starting point for anybody new to the field [1,2,3,4,5]. The term "invention" in our title refers to the fact that we will here concentrate on some specific "development/ discover" aspects for (planar or in patterned substrates) which, generally, are not thoroughly discussed in current research papers (or to say better, not necessarily in the context of the same review), partly because these have been the result of historical meanderings of which it is, indeed, difficult to give a succinct account. Considering that current endeavors are largely run by, essentially, second and third generation researchers, we hope that our effort will be useful for those early stage researchers which want to get a more in-depth knowledge in the early field developments.

We consider, anyway, acknowledging these aspects important for a proper understanding of the QD field, some of its idiosyncrasies and, somehow, its edges for future directions. For this reason, we will present a, fast-tracked and necessarily incomplete historical overview concentrating only on some specific material and quantum dot formation/epitaxy systems, with somehow a focus on single dot growth developments, more than dealing in detail with all the advances.

It will become also clear that QD research has largely progressed as a community endeavor over the years, and that many conceptual developments owe a lot from very different fields of physics and technology, from surface science to semiconductor growth and, being somehow application driven, quantum transport and quantum optics.

## 2. The beginning

The general expectation that "small" semiconductor confined structures would be of strong interests in terms of their physical properties and potential applications followed closely the technological developments that made possible to actually "build" such structures. While the general understanding that confined structures will present discrete energy level was deeply rooted in quantum mechanics from its inception, one has to wait the sixties and even more the seventies to see the beginning of significant literature on the importance of applications of quantum confinement (in general referred to as quantum size effects in the early stages, QSE), with the quantum well (QW) laser invention being one notable result.[6,7,8,9] We observe in this context that one could still publish papers of interest to the community in the seventies on "*the quantum-mechanical problem of the simple harmonic oscillator confined into a box has been solved*"[10], showing that there still were specific problems which needed addressing. It is worth noting also that while 1D confined structures (as we refer to them today) quickly acquired the quantum well labelling (not immediately though, for example the, arguably, first QW laser (1977) for today's standards by Rezek et al. in ref 11, does not use the quantum well terminology, used nevertheless quickly after e.g., in references 12,13), one has to wait the nineties for a broadly agreed labelling of lower dimensionality structures (see also our discussion below).

It should be strongly underlined that applications (both in terms of fundamental physics[14] and devices, see e.g. ref. 15) were also made possible by relevant advances in (semiconductor) technologies[16]. It is worth reminding that molecular beam epitaxy (MBE) and metalorganic vapour phase epitaxy (also known as metalorganic chemical vapour deposition and variation of thereof, MOVPE, MOCVD, OMVPE, OMCVD, CVD), saw their consecration during the seventies, become available to several groups as industrial products during the early eighties[17,18,19,20,21,22] and replaced other "growth" techniques which were deemed less versatile. It is also important to note, that the invention of the atomic force microscope-AFM (somehow closely following the invention of the Scanning tunneling microscope-STM)[23] was industrially made available to several research groups only in the early nineties. Indeed, in many early publications in the field, electron microscopy in the form of TEM (transmission electron microscopy) and SEM (scanning/secondary electron microscopy) were the main source of microscopy structural feedback (and to be added as well, in situ diffraction techniques such as low energy electron diffraction, LEED, and reflection high energy electron diffraction, RHEED). Adding to this argument, many of the recent results on "quantum light" have been made possible by the broader availability of more advanced (and sometimes in some way cheaper) laser sources, one for all the (mode-locked) Ti-Sapphire laser in the late eighties and early nineties, and, more recently, also thanks to significantly improved single photon detectors, based on various technologies.[24]

## 3. The first proposals for quantum dot applications

The manuscript by Arakawa et al. in (ref. 25 [1982], "*Multidimensional quantum well laser and temperature dependence of its threshold current*"), is (correctly so) commonly cited as a landmark paper for the development of quantum dot technologies. Indeed, it has been a strongly motivational theoretical analysis suggesting that quantum dot lasers (or better, low dimensionality ones beyond QWs) should help delivering more efficient and more temperature stable laser structures. The manuscript appeared just before such structures could be obtained with semiconductors in any controllable fashion, and is well known to contain a critical approximation in its derivation: i.e. it assumes that the quantum dot structures can be obtained uniformly identical across the active layer (citing and extrapolating from the paper: "*In 3D-QW lasers, the thermal spreading of carriers should vanish because the state density is delta-function like. Hence, the temperature dependence of $J_{th}$ will totally disappear, as long as the electron population in higher subbands remains negligibly small*"). While quantum dot lasers are a (commercial) reality today[26,27,28] and indeed generally show a better temperature response, it should be noted that, nevertheless, the quantum dot formation comes with a distribution of sizes, as self-assembled processes are used. This affects the maximum performances attainable, a limit which was indeed promptly underlined already in the early stages of QD research[29].

Here we are also interested in highlighting the choice of terminology: quantum dot structures are not called, in ref 25, zero dimensional (0D) structures, but are referred to as 3D-QW (a quantum well with a 3D confinement, and a quantum wire is a quantum well with a 2D dimensional confinement in the manuscript), showing that still there is no consensus in 1982 on how to label nanostructures. During the eighties and the early nineties it is indeed not unusual to find notations talking about QW wires and boxes (and clusters), which then evolved into quantum islands, boxes, dots, and finally converged to the current practice. The interested reader should then pay attention during (electronic) manuscript content searches to cover all the possible nomenclatures when interested in the early literature.

It is also worth noticing that the authors of ref 25 did not actually expect (in their words)[30] epitaxy to progress so rapidly to deliver quantum dot structures in a short time, and so they cleverly proposed and tested the use of magnetic field as an alternative tool to increase (reduce) confinement (dimensionality): "*In view of the technical difficulties of fabricating such structures, we have investigated the conventional GaAs-GaAlAs DH laser placed in a strong magnetic field*".

## 4. Top-down vs bottom-up

Despite the pessimism of some, nanostructures (as we would call them today) developments appeared straight away and quickly, on parallel strands. On one hand, the scientific (semiconductor) community addressed confined structures by top-down approaches, i.e. patterning semiconductors to achieve low dimensional structures, micron or nm sized. We cite here only some representative manuscripts on this topic for the interested reader, and underline that the community seeking optical effects quickly realized, during the eighties, that top-down approaches (including implantation) could bring with them difficult to eliminate distributed defects affecting widespread good optical features, uniformity and control.[31,32,33,34,35,36,37,38,39] This has been one of the main drive ("*There is growing, universal evidence that one-dimensional (1D) and OD nanostructures lead to diminished luminescence quality when compared to 2D or 3D heterostructures*") [40] to search for bottom up self-assembled structures, which showed stronger promises once tested, and, had also effectively begun in the early eighties, sometimes through chance findings as we will discuss.

It should be underlined that the electrical transport community persevered on the top-down approach, obtaining significantly good results. They eventually evolved in time towards lithographically defined structures which limited etching steps and with the low dimensional confinement obtained largely with simple metal contacts and electrical field defined geometries, and we cite here in references some non-exhaustive examples for the interested reader of both categories.[41,42,43,44,45,46,47,48] Self-assembling in the transport community, nonetheless, has been / is the subject of important research lines, as for example the search for Majorana states with semiconductor nanowire structures[49].

## 5. Early observations of self-assembling, the origins of the Stranski–Krastanov (SK) terminology and the first established results also in other systems

Before semiconductor self-assembling was experimentally observed on semiconductor planar structures, since the sixties surface scientists had started depositing by evaporation monolayers (MLs) of material A on crystalline material B and explored the growth (ordered/epitaxial) processes involved. Researchers focused for example on metal on metal and metals on semiconductors, and then quickly broadened their interests into more exotic compounds, relying for interpretation often simply on photoemission spectroscopy and interference technologies, such as LEED. As early as 1958[50] one finds a (theoretical) classification of growth modes akin to the ones still in use today: "*Of the three different growth mechanisms, that of growth atom-layer by atom-layer through surface nucleation (FRANK-VAN DER MERWE mechanism) occurs only for growth on material of the same composition. The growth on a different material, on the other hand, occurs, in general, by original formation of a mono-or poly-molecular layer without nucleation, which then leads to formation of three-dimensional nuclei (STRANSKI-KRASTANOV mechanism), or through the formation of three dimensional nuclei directly on the foreign substrate (VOLMER-WEBER mechanism)*".

While today we would not consider the substrate of choice as so relevant for the epitaxial growth mode description (we certainly do not consider a layer by layer growth as happening only if the epitaxial material is the same as the substrate), it is worth noting that this growth mode discussion appears first and mostly as "in German" literature (i.e. many of the originals were written in German, some works referring to manuscript which appeared as early as the twenties/thirties, not in general referring to pure epitaxial semiconductor processing as referred today[51], but ordered processes in crystal precipitation for example, as it is expected, in view of the historical period). Only subsequently, the classification would be adopted by the whole community. Here we report, also as references and for completeness, the manuscripts normally cited as the source of the classification (not all in the original ref. 50 actually, and we caution the author that those references are not always straightforward and revealing in the context of the mentioned classification, which is largely to be attributed as being an original compendium from E. Bauer and his work in ref 50).[52,53,54]

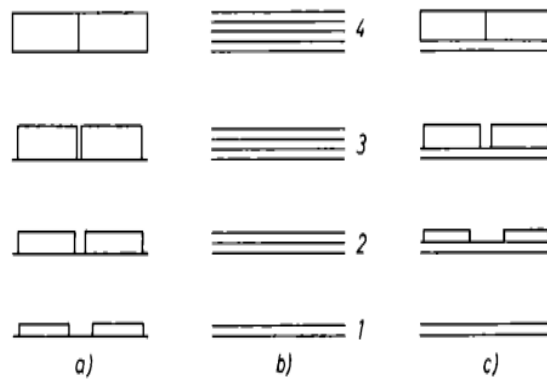

Abb. 3. Die verschiedenen Wachstumsmechanismen:
a) VOLMER-WEBER-Mechanismus, b) FRANK-VAN-DER-MERWE-Mechanismus,
c) STRANSKI-KRASTANOV-Mechanismus

*Fig. 1. The original schematic of the three growth modes as in ref. 50, including the original "in German" figure caption*

The classification in ref. 50 did not have an immediate widespread use, with only few authors addressing it outside the German community at first.[55] Nevertheless, in the early seventies it started to be largely used by the developing and growing surface science community, and by the end of the seventies it became the dominant description utilized in publications. As far as this review is concerned, the Stranski-Krastanov classification is one of particular interest, as today it is, notably, almost only used in relation to quantum dots. Volmer–Weber quantum dots (the mechanism is not conceptually too different from a droplet of liquid aggregating and not wetting a surface), it should be said, are also possible and have been indeed been reported later on [56,57], but the QD success story is unquestionably linked to SK dots, and deserves a broader discussion.

A first note of interest is associated to the fact that the original Stranski-Krastanov growth mode description somehow deviates substantially from the dominant current utilization. The case discussed in ref. 50 citing "Stranski –Krastanov" processes in ref. 53 refers (summarizing) to *the growth of an ionic AB crystal with the ionic charge of ±1 on an AB "substrate" crystal with the ionic charge of ±2. In the discussion, this has implications for the binding energy of the first (new) lattice plane and, as a consequence, on the vapor pressure of the ionic compound, which changes for the second lattice plane due to electrostatic repulsion. In that picture, a supersaturation of the ionic compound is necessary for growth and overall, electrostatic interaction/binding energies and vapor pressures will determine, where and under which conditions growth takes place. The deposition of a crystal lattice plane will not heal any defects or produce growth within the upper lattice planes of the substrate crystal. Any steps and lattice defects in the "substrate crystal" will lead to enhanced crystal (3D) growth. Growth is especially initiated at non-finished layer parts of the substrate crystal at edges and corners*.

Indeed, the original discussion related mostly to a layer by layer growth which because of impurities or dislocations would then decide to turn 3D, forming islands as a form of energy minimization. In the original manuscript, the breach of the crystal perfection was basically (nearly) intrinsic to the description.

Since the classification of ref. 50, a few-layer-growth followed by 3D growth is labelled as SK, whatever the specific driving mechanism[58]. It should nevertheless be noted that, today, defect induced 3D nucleation is not generally described as a Stranski-Krastanov process, or at least, and more precisely, not discussed as such when it is not useful for future developments. The 3D formation, when semiconductor QDs are concerned, is nowadays mostly only linked to a specific energy minimization mechanism, i.e. that 3D growth reduces the strain energy for lattice mismatched growth (basically thanks to lateral relaxation), a process that supersedes the increase in surface energy of the system. This evolution in the specific meaning and usage of the terminology is indeed peculiar, and is worth underlining. Indeed, a more "mechanism" specific growth mode classification[59] could have been possible historically (but it did not happen). So, when addressing the literature in the seventies and eighties, it is important to note that SK processes covered a broader range of nucleation mechanisms then today's conventional description associated to QDs.

Nomenclature aside, during the eighties self-assembling processes were observed and somehow sought after by the community, partly because of inevitable surface science links. Nevertheless, truly *controlled* self-assembly for semiconductors is reported first in the early nineties (and the word "self-assembled" itself appears nearly at the same time[60] to be broadly used in the literature in relation to quantum dots or quantum wires). It is nevertheless interesting to discuss a few early results which would constitute the foundations of the subsequent developments.

"Self-assembling" concepts appear in the literature of the eighties in association to different semiconductor epitaxial processes, and different deposition techniques. For example, ref. 31 observes in 1982 wire like emission (there called quantum-well wires) for GaAs/AlGaAs structures as a result of a combination of epitaxial growth, lithography and regrowth. Quantum-well wire structures were interesting for laser applications[61], and alternative approaches were also investigated. Progress for self-assembling is fast as soon as several authors start exploring complex growth structures: for example in 1987 ref. 62 reports on AlAs/GaAs superlattices on misoriented surfaces and concludes: "*From these results, the lateral controlled MOCVD technique appears very promising to provide quantum well wires with dimensions less than 100 X 100 Å$^2$ without resorting to lithographic processes*". Growth on patterned substrates also allowed to observe (self-assembled) quantum wire behaviors.[63] Pre-patterning had shown promises two years before[64], as the crystallographic facet dependent growth rates in MOVPE would be the precursor to site-controlled quantum wires and quantum dots self-assembling.[65] In 1988 reference 66 reports in the abstract "*A unique feature of this laser is the formation of a quantum-wire-like crescent shaped active region at the center of a two-dimensional optical waveguide.*" These efforts also inspired the silicon community, which also explored self-assembling as tools to reduce dimensionality. In 1993 reference 67 reports on the "*Realization of crescent-shaped SiGe quantum wire structures*". Soon, further dimensionality reduction would be achieved with similar methodologies, and 3D site-controlled quantum dot-like structures in the III-V system would be first reported in 1995[68]. It is also interesting to observe that some of the discussed results and ideas can be traced back to more general concepts of self-assembling, developed 20-30 years before. For example, in reference 69, the author discusses, theoretically, surface evolution and "*developments described by the formulas…(which) are idealized representations of the latter stages of the sintering of small wires and particles to a plane.*"

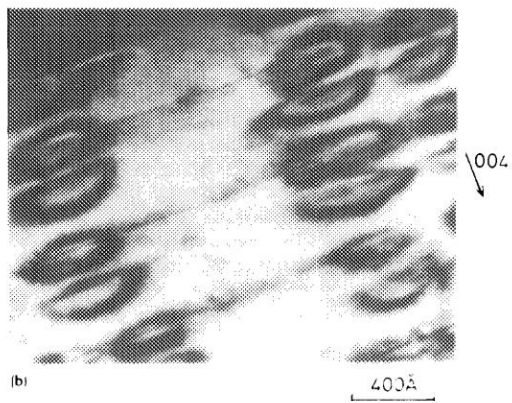

*Fig. 2. Reprinted from ref. 71, Scanning transmission electron microscopy (STEM) showing evidence of 3D nucleation in InAs/GaAs superlattices. Note: the SK terminology is not present in the manuscript.*

All these developments followed also a rich pattern of sometimes "serendipitous" observations during the eighties and early 90s that would contribute to positive anticipations in respect to obtaining self-assembled nanostructures. Various early paper have been identified in the literature as seminal for SK QD developments. Without entering the debate, it is clear that while the surface science community led the concept, the III-V community also first hit into it through a different path. In general, discoveries were linked to In(Ga)As deposition that had been investigated to figure out relaxation thresholds in strained layers, to look for strategies for abrupt interfaces in superlattice structures etc.. It is also worth underlining that early literature on SK processes, did not necessarily have the experimental tools to distinguish which mode of SK they were observing, i.e. if the one following defect formation (in general strain relaxation through dislocation formation) or they obtained "coherent" defect free dots. As will be apparent in the following discussion, SK QDs developed as a useful semiconductor technology, only after the community convinced itself that it was possible to obtain defect free structures.

In the process of optimizing 2D epi-growth protocols, the community hit into 3D structures, clustering etc. linked to strain relaxation generally associated to defect formation. As early as 1983 one can read in one abstract[70]: "*Results of investigations by reflection electron diffraction show that the nucleation of InAs/GaAs (100) occurs via island formation and coalescence*". In the same manuscript text, following RHEED observations in reciprocal space from which one can infer lattice parameters: "*We observe that at an equivalent coverage of 7 Å of InAs…the entire pattern remains sharp and gives the same lattice constant as GaAs(100)… The diffraction pattern observed at 9 Å coverage, however, is markedly different, In addition to the pattern previously recorded, another complete pattern is superimposed with the same symmetry and relative intensity distribution but slightly larger lattice constant*", and later in the conclusions "*(3) the nucleation of InAs/GaAs (100) is via island formation and coalescence, (4) these islands of InAs are not coherently constrained to the underlying GaAs substrate lattice constant*". The basics of an SK process (especially if one considers the version where defect formation are intrinsic to the process) are somehow already spelt out, even if the specific focus was not on self-assembling, and there is no broader discussion, e.g. on the threshold for defect formation (growth condition related). In 1985 ref. 71 (and soon others around the same years) adds on those results and discusses that "*It has been shown that even when In-rich clusters are formed, good crystalline quality material can be obtained. Also, specific and intense photoluminescence is associated with the cluster formation. These kinds of structures are thus proved to be of interest to study low-dimensional ( < 2) objects showing good optical properties.*" In ref. 72 (1987) appears the phrase "*Our data reveal a Stranski-Krastanov growth mode, i.e. the formation of islands on a uniform layer.*" Other reports in those years and by different groups observe islanding, not always attributing those self-assembling to the Stranski-Krastanov growth mode, which, in this early literature, is often mixed to more cautious vocabulary (e.g. cluster, islands etc.) and across other semiconductor material systems too. It is maybe also worth mentioning the (1991) discussion in ref. 73, highlighting that the developed explanation about cluster relaxation through defect formation had to be revised, as microscopy observations showed a number of relaxed islands not associated to defects ("*In contrast to traditional critical-thickness theories, significant strain relief is accommodated by a coherent island morphology. This study represents a new view for both the growth mode and initial strain relaxation in thin films*").[74] Also several other papers while discussing the observation of clustering, islanding and/or Stranski-Krastanov processes, actually aimed to reduce 3D effects in 2D epitaxy, and

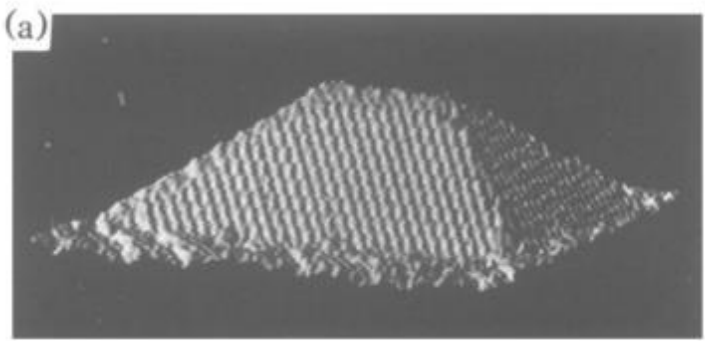

*Fig. 3. Early results in the SiGe system: from ref. 78 (1991), documenting the STM images of a single "hut" cluster, as it is called in the manuscript.*

while doing that also contributed to the understanding of self-assembling. It is worth mentioning those contributions too. The interested reader will find in the following references an attempt to represent the broadness of the epitaxial effort involved (which happened across continents, and across platforms, see for example the Si/Ge research), which built the necessary understanding for a number of important outcomes which were to immediately follow.[75,76,77,78,79,80,81,82,83]

Indeed, the establishment of SK dots as a powerhouse for low dimensional physics sees it institution in the early nineties with several manuscript (actually a real "boom" of) across just a few years presenting important results, with, in general, widespread claims of dislocation free coherent island formation. [84,85,86,87,88,89,90,91,92,93,94,95,96,97,98]

By inspection of the listed references, it is clear that between 1992 and 1994 a cornucopia of results appear under the different form of SK QD labelling and a number of different groups independently develop their technological processes, all motivated by the scientific opportunities of low dimensional structures. Notably, the early observations of reasonably "good" photoluminescence from ensembles already reported in ref. 71 and confirmed in later publications (see e.g. reference 87, somehow representative of a number of publications which appeared in 1994 and relevant as incipit for a large number of future other publications) was followed by the "single" dot observations (1994), e.g. in ref. 94 ("*and observe for the first time the emission of single such quantum dots*"). This is paralleled by the first proposals for electrically driven devices capable of single photon emission.[99]

Stranski-Krastanov dots are not obviously the only self-assembled QDs which entered the arena[100], aside of the already mentioned site-controlled dots. The early nineties have seen a wealth of results and alternative methods to develop low dimensional structures. For example, reference 101 (1993), building on previous works, reports on the realization of "*GaAs pinched-off pyramidal volumes of base - 50 nm and height 13 nm*", discussing structures which are processed to stand-out of the semiconductor substrate, and started a full line of research in that direction. In ref 102 (1996) interface GaAs QD in AlGaAs barrier (i.e. relatively weakly confined QDs resulting from monolayer level interface disorder) were discussed and would represent in the following years a relevant system for testing fundamental quantum optics properties (see e.g. ref. 103).

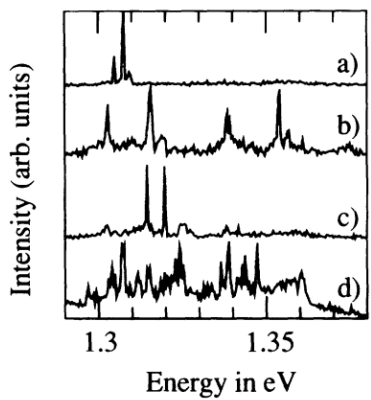

*Fig. 4. Single QD photoluminescence from reference 94*

A nowadays well-established well exploited alternative for obtaining quantum dots (today named droplet epitaxy) also appeared around 1991 ("*New MBE growth method for InSb quantum well boxes.*") in reference 104, and was demonstrated on the exotic InSb on CdTe ("*Then 200 nm thick CdTe epitaxial layer was grown onto the InSb buffer layer at 200°C for 30 mn. Next, In(dium), which had the same flux during growth of the InSb buffer layer, was deposited on the CdTe epitaxial layer for 30s at 200°C. After deposition of the In droplets, an Sb flux which was 1 x $10^{14}$ atoms/$cm^2 \cdot s$ was supplied for 100s*"), following earlier results (1990) were the idea was first spelt out, see for example Ref. 105 dealing with "*GaAs micro crystals.. using Ga droplets*" on ZnSe. The first GaAs on AlGaAs version we are more used today also quickly followed.[106]

It is also interesting that those "droplet" results touched a bit towards other processes explored by a different research community and somehow considered as a separate technology: in reference 107 ("*Growth of GaAs Epitaxial Microcrystals on a S-terminated GaAs (001) by VLS Mechanism in MBE*") an initial spurious vapour liquid solid (VLS) process is (probably) observed and reported, an outcome which was accidentally witnessed during the investigation of droplet epitaxy growth protocols. Reference 98 cites the 1964 paper[108] reporting on the proposal/interpretation of a VLS process for silicon whiskers growth. It should be underlined that semiconductor whiskers (now referred often as nanowires) had been investigated constantly during all those years, and an early account with III-V materials can be found for example in reference 109. Nowadays there is broad community working on different topics related to VLS nanowires, including embedding a single or multiple thin QD layers, and we limit ourselves to cite here only a few recent review manuscript for the interested reader.[110,111]

The epitaxial results obviously continued hand in hand to the theoretical understanding of the processes involved, which overlaps with the parallel surface science findings. These also built on the early theoretical findings we discussed earlier in this manuscript. For example, a more fundamental understanding in the semiconductor community for the SK dot system (and the one less intuitive to explain in its "defect free" version) of the combined role of surface strain and relaxation to achieve coherent islanding nevertheless was relatively slow to develop at first. In ref. 84 (1990) one can still read, about SiGe islanding, "*we therefore propose that elastic deformation of the substrate and the island lowers the strain energy of islands (under all circumstances) and that under certain conditions this strain relaxation will be sufficient to allow coherent Stranski-Krastanow island formation to occur*", and "*A suggestion that this latter process may be occurring can be found in the heavy strain contrast at the islands in Fig. 2, suggesting significant local strain relaxation near an island. A strain field such as that shown schematically in Fig. 3 would relax misfit strain in the island, thereby allowing islanding without dislocation introduction (even under SK conditions, i.e., where interface energies alone would dictate layer-by-layer growth)*", showing a fundamentally correct description, but, for example, not explicitly mentioning the competing surface energy contribution. That said, the surface science community had already developed a more profound understanding, as is evident for example from reference 73 (1991), which in the process of discussing the 3D islanding as being closer to Wolmer Weber than Stranski-Krastanov in their cases, mentions: "*because

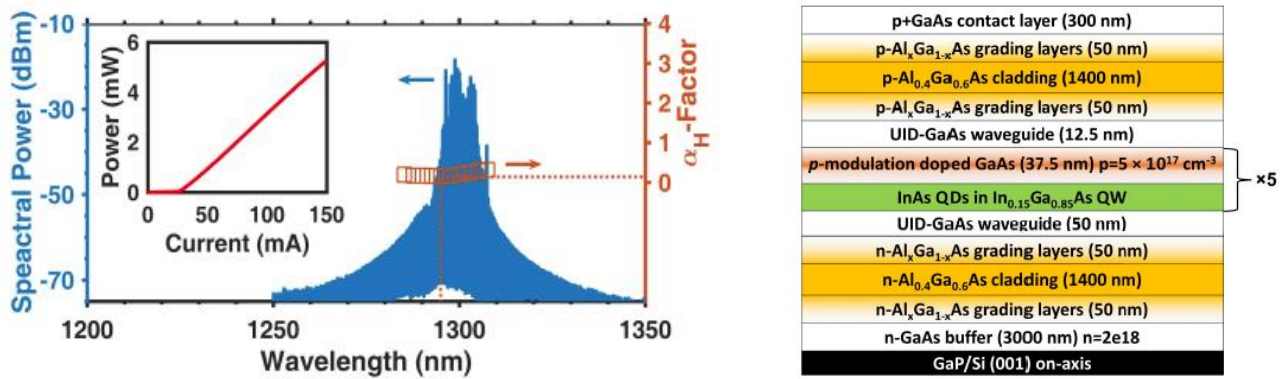

*Fig.5. Left: photoluminescence of a QD laser epitaxially grown on Silicon from reference 122. The device exhibits a very large ES-to-GS lasing threshold ratio, which is a figure of merit for a high degree of stability. On the right the complete sample structure description.*

*the critical cluster size, for which the reduction in strain energy dominates the additional surface energy, is too large*" citing in the references a difficult to find manuscript itself, but whose content should be well represented by ref. 112 (1988), which is fundamentally giving a substantially correct and complete depiction of the energetics of current QD SK processes.

Nevertheless, in just a few years a more complete widespread theoretical understanding quickly developed[113,114,115,116] and by the second half of the nineties theoretical works were already concentrated on addressing specialized issues (for example one reads in ref. 117 "*In this paper, we present a true kinetic model that yields the time dependence of the 3D island density and describes how this density varies with external growth conditions*"). This relevantly includes a more detailed analysis of the wetting layer role (and its dependence on growth conditions), and its important influence on the overall dot formation processes.[118,119]

It should be observed that for other less popular QD systems but equally not easy to describe theoretically, as for example the so-called pyramidal quantum dot system, one has to wait the first decade of this millennium for a thorough theoretical description, partly because of its reliance on MOVPE-only specific properties.[120,121]

### 6. Perspective

What we discussed above, has been a very successful and rapid journey in establishing QD technology. And, even if QD research has now evolved towards a "mature" status, there is still ample space for new developments and applications. For example, the initial drive (quantum dot laser) did produce commercial products presently competing on the telecom market, and, nowadays, also bran new developments are researched. On one front because the higher temperature stability is a strong plus in a future of photonic integration and reduced power consumption in data centres (i.e. less cooling needs) is a relevant motivation. Reduced sensitivity to reflections[122], lower linewidth enhancement factor and reduced linewidth are also favorable properties in an integration perspective. On the other hand, SK dot lasers have also recently showed a path for low cost integration on silicon wafers, with SK dot layers acting as both dislocation filters to intercept relaxation defects from the III-V material epitaxially deposited on silicon, and as "sturdy" active layers in the same structures, as, in a simplified picture, single defects and dislocation will only affect single dots, and not the whole layer as in a QW laser.[28,123,] Other captivating applications also enlist QDs as periodic strain relief tools.[124]

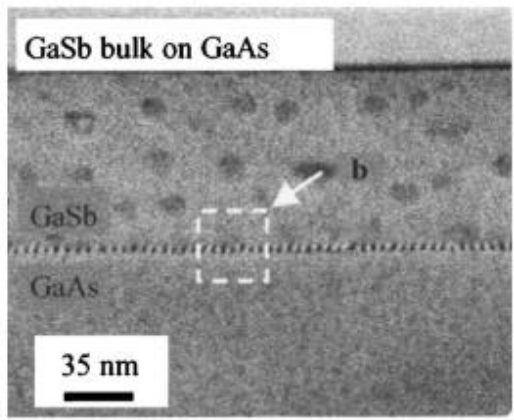

*Fig.6 from ref. 124. Periodic array of misfit dislocations at the GaSb layer GaAs substrate interface (induced by early stage islanding).*

Moreover, since the early 2000 quantum dots have become the centre of an intense quantum technology effort, as source of quantum light. The potentially disputed nature of "artificial atoms" was repeatedly confirmed in several avenues, and citing just a few, major breakthroughs were single photon[125] and entangled photon emission[126], including high throughput of identical photons[127] also thanks to bespoke cavities[128]. Those results are important stepping stones towards applications, and between the next years challenges cluster state generation[129] and associated engineered quantum coupling are probably the ones bearing the strongest potential for impact[130]. On the other hand quantum dots (and quantum wires) are far from having exhausted the fundamental research opportunities, and more exotic results should be expected in the next few years, for example in the field of Wigner molecules for topological quantum computation[131] and/or Majorana quasi particles.[132]

All these topics indeed face the same challenges, i.e. evolving the well-characterized structures to new bespoke ones, improving control on the epitaxial front, evolving device processing and taming advanced characterization and compatibility with hetero-integration.[133]

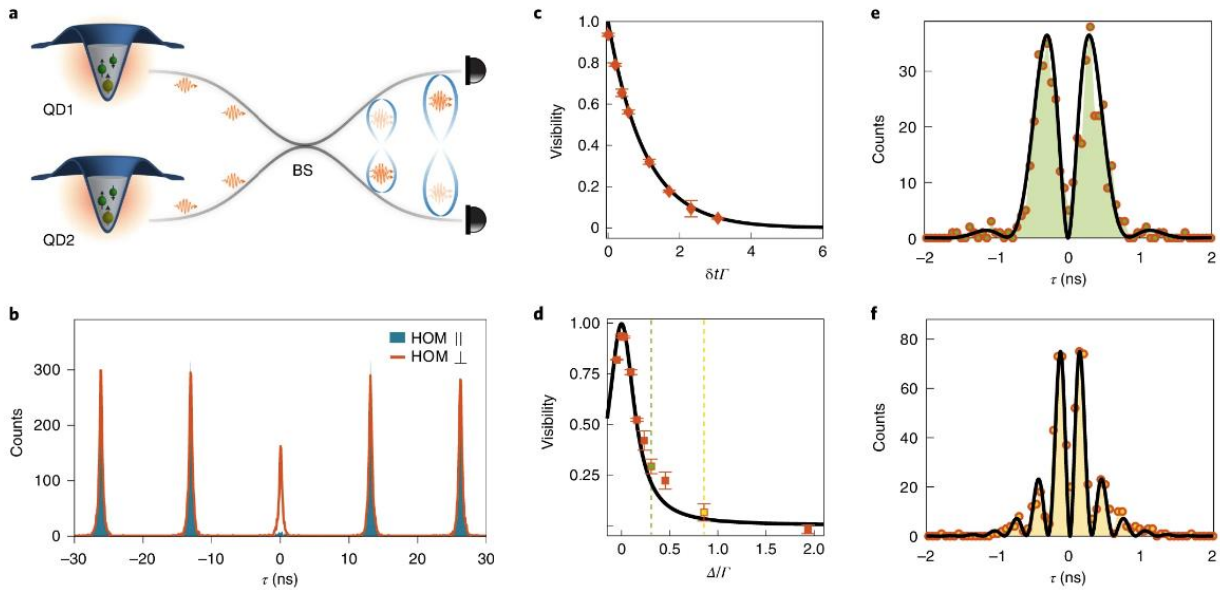

*Fig.7. from ref. 127 showing quantum interference from identical photons. For simplicity, we report also the original figure caption below.*

*a, Schematic of quantum interference between photons from two remote GaAs QDs. Provided that the photons are indistinguishable, they coalesce and exit the beamsplitter (BS) at the same port. b, HOM interference from photons generated by two remote QDs. The vanishing central peak shows a raw HOM visibility of (90.9 ± 0.8)% and a true two-photon interference visibility of (93.0 ± 0.8)%. c,d, Two-QD HOM visibility as a function of normalized temporal delay δt and spectral detuning Δ between the two QDs. The QD2 photons are delayed with respect to QD1 by delay δt (c). Frequency detuning Δ between the two QDs is varied by exploiting the exquisite frequency control provided by the electrical gates (d). The uncertainties in interference visibilities in b–d represent the 1σ error. e,f, Pronounced quantum beats are revealed in the two-QD HOM central peak when the QDs are slightly detuned in frequency. The green (e) and yellow (f) colours indicate the detuning in d. The solid lines in c–f show the quantum–optical theory. Here we assume the wavepackets of remote QDs are identical except for a small delay in c and small detuning in d–f.*

## Acknowledgments

We thanks Armando Rastelli for the motivating discussions on SK dot early stage categorization, and Francesco Montalenti for his useful comments on the manuscript and his insights on the theoretical aspects.

This research was supported by Science Foundation Ireland under Grant No. 15/IA/2864 and 12/RC/2276_P2.